\documentclass[prb,twocolumn,superscriptaddress,showpacs]{revtex4} 
\usepackage[dvips]{graphicx} 
\usepackage{latexsym} 
\usepackage{bm} 
\usepackage{psfrag} 
\usepackage{wasysym} 
\usepackage{colordvi} 
 
\begin{document} 
 
\author{D. N. Sheng} 
\affiliation{Department of Physics and Astronomy, California State University, 
Northridge, CA 91330} 
\author{L. Sheng} 
\affiliation{Department of Physics and Texas Center for Superconductivity, University of 
Houston, Houston, TX 77204} 
\author{Z. Y. Weng} 
\affiliation{Center for Advanced Study, Tsinghua University, Beijing 100084, China} 
\author{F. D. M. Haldane} 
\affiliation{Department of Physics, Princeton University, Princeton, NJ 08544} 
\title{ Spin Hall Effect and Spin Transfer in Disordered Rashba Model } 
 
\begin{abstract} 
Based on numerical study of the Rashba model, we show that the spin Hall 
conductance remains finite in the presence of disorder up to a 
characteristic length scale, beyond which it vanishes exponentially with the 
system size. We further perform a Laughlin's gauge experiment numerically 
and find that all energy levels cannot cross each other during an adiabatic 
insertion of the flux in accordance with the general level-repulsion rule. 
It results in zero spin transfer between two edges of the sample as each 
state always evolves back after the insertion of one flux quantum, in 
contrast to the quantum Hall effect. It implies that the intrinsic spin Hall 
effect vanishes with the turn-on of disorder. 
 
\typeout{polish abstract} 
\end{abstract} 
 
\pacs{72.15.-v, 71.30+h, 72.10.-d} 
\maketitle 
 
The two dimensional (2D) electron systems exhibit integer and fractional 
quantum Hall effect (IQHE/FQHE) in the presence of strong magnetic field~% 
\cite{qheb}. The precise connections between the exact quantization of Hall 
conductance, the topological property of the wavefunctions, and the current 
carrying edge states in open systems have been well established~\cite% 
{qheb,thou,laugh,halp,arou,top}. As the hallmark of a QHE, the quantized 
plateaus~\cite{qheb} appear in the presence of disorder where the 
longitudinal conductance vanishes with the opening of a mobility gap, which 
ensures a robust dissipationless transport regime. In an open system, the 
edge states are the chiral current carrying states free from backward 
scattering, which leads to a precise quantized charge transfer between edges 
of the sample as demonstrated through Laughlin's gauge argument \cite% 
{laugh,halp,top}. 
 
The recent proposals of the intrinsic and dissipationless spin Hall effect 
(SHE) in three-dimensional $p$-doped semiconductors~\cite{zhang} and 2D 
electron systems with Rashba spin-orbital coupling (SOC) ~\cite{sinova} have 
stimulated a great deal of interest~\cite{sh2,sh4,sh7, 
sh8,rashba,sh9,sh10,sp3,sp4,sp5,chen}. Such intrinsic SHE may provide a new 
way to manipulate electron spins in nonmagnetic semiconductors without the 
application of magnetic fields. It was also suggested that the recent 
experimentally observed spin polarization or accumulation~\cite{sp3} in 
electron systems might be related to this effect~\cite{sp4,sp5}. 
 
However, there exits a strong debate regarding the fate of SHE in the 
presence of disorder. A challenge to SHE in the Rashba model comes from 
analytical perturbative calculations involving vertex corrections, where SHE 
is found to vanish in the presence of any weak disorder ~\cite{sh7,sh8}. 
On the other hand, numerical calculations based on the Kubo formula\cite{sh9} 
using continuum model in momentum space found that spin Hall conductance 
(SHC) is finite for finite-size system with weak disorder, while its proper 
thermodynamic limit is not clear. 
%%The calculations based on the Landauer-B% 
%%\"{u}ttiker (LB) formula with four semi-infinite leads further suggested~% 
%%\cite{sh10} that SHE could be robust against rather strong disorder 
%%scattering, where the boundary effect from the attached leads may play an 
%%important role. 
A consistent picture for the SHE is still absent. Especially 
the topological property of the SHE has not been well addressed, while all 
conventional quantum Hall systems are considered to have nontrivial 
topological origin\cite{arou,top}, including the Hall effect without 
magnetic field\cite{haldane,haldane1}. 
 
In this paper, we present a numerical Kubo formula calculation of 
SHC based on the Rashba lattice model. We show that the finite SHC 
persists in the presence of random disorder scattering for 
finite-size systems up to a characteristic length scale, beyond 
which SHC decreases and vanishes exponentially. The scaling 
behavior of the SHC follows a one-parameter scaling law in a 
fashion similar to the longitudinal conductance in a localized 
system~\cite{local}. Furthermore, we perform numerically a 
Laughlin's gauge experiment\cite{laugh,halp} by adiabatically 
inserting flux to directly probe the experimentally measurable 
spin transfer and accumulation associated with SHE. We find that 
energy levels do not cross each other during the adiabatic 
insertion of the flux once the disorder is turned on, in 
accordance with the level-repulsion rule. It results in zero spin 
transfer between the edges and zero spin accumulation at the 
boundary, as each state always evolves back to itself  after the 
insertion of one flux 
quantum. This is in contrast to the quantized charge transfer in the IQHE% 
\cite{laugh,halp}, which is associated with the nontrivial topological 
property of quantum Hall systems\cite{arou,top}. We thus conclude that the 
topological SHE vanishes in the disordered 2D Rashba model. An example of a 
topological SHE in 2D electron system\cite{haldane1,kane} is also briefly 
discussed. 
 
We start with a tight-binding lattice model of noninteracting electrons with 
the Rashba SOC. The Hamiltonian is given as follows:~\cite{ando,dns96} 
\begin{eqnarray}  \label{soc} 
H&=&-t\sum_{\langle ij\rangle} \hat c_{i}^\dagger \hat c_{j}+\sum_{i} w_{i} 
\hat c_{i}^\dagger \hat c_{i}  \nonumber \\ 
&+&V_{so}\left(i\sum_{i }\hat{c}_i^+\sigma _x \hat{c}_{i+\hat{y}}-i\sum_{i }% 
\hat{c}_i^+\sigma _y\hat{c}_{i+\hat {x}} +H.c.\right) \ .  \label{ham} 
\end{eqnarray} 
Here, $\hat{c}_i^+=(c^+_{i\uparrow},c^+_{i\downarrow})$ are electron 
creation operators with $t$ as the nearest-neighbor hopping integral and $% 
V_{so}$ the Rashba SOC strength, and $w_i$ is an on-site random nonmagnetic 
potential uniformly distributed in the interval ($-W/2, W/2$). 
 
We numerically study square samples of side $L$ by using exact 
diagonalization method under a twisted boundary condition: $\hat{\Psi }(i+L% 
\hat{x})=e^{i \theta_x}\hat {\Psi} (i)$ and $\hat{\Psi }(i+L\hat{y})=e^{i 
\theta_y}\hat {\Psi} (i)$, where the electron wave function, $\hat{\Psi}% 
(i)=\left( 
\begin{array}{c} 
\Psi_{\uparrow} (i) \\ 
\Psi_{\downarrow} (i)% 
\end{array}% 
\right )$, has two spin components. The linear response SHC at zero 
temperature can be calculated by using the Kubo formula 
\begin{eqnarray}  \label{kubo} 
\sigma_{sH}=\frac {-e\hbar }{N} \sum_{E_m < E_f<E_n} \frac {\mbox{Im}% 
\left(\langle \hat{\Psi}_m|J^{z spin}_x|\hat{\Psi}_n\rangle\langle \hat{\Psi}% 
_n|v_y|\hat{\Psi}_m\rangle\right)} {(E_n-E_m )^2 } \   \label{kubo} 
\end{eqnarray} 
where $E_f$ is the electron Fermi energy, and $\hat{\Psi}_m$ represents the $% 
m$-th eigenstate with energy $E_m$. We use the standard velocity operator $% 
\mathbf{v}=\frac i {\hbar } [H, \mathbf{r}]$ ($\mathbf{r}$ is the position 
operator of electron) and the spin current operator $J^ {z spin}_x = \frac {% 
\hbar} 4 \{v_x, \sigma_z\}$, which measures spin current flowing along the $x 
$ direction with the spin polarization along the $z$ direction. 
 
\begin{figure}[tbp] 
\begin{center} 
\vskip-2.8cm \hspace*{0.0cm} \includegraphics[width=2.2in]{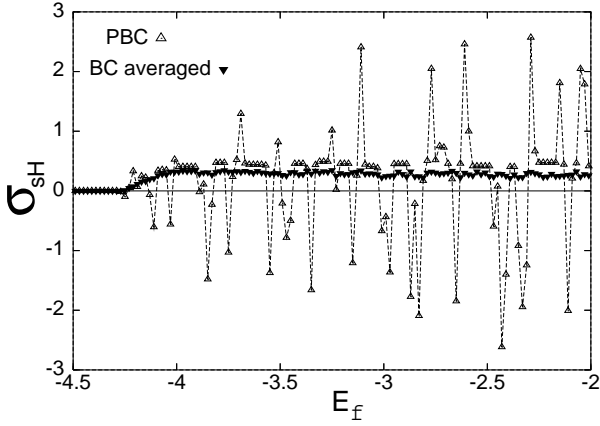} \vskip% 
-2mm \vskip-8mm 
\end{center} 
\caption{Spin Hall conductance $\protect\sigma_{sH}$ (in units of $e/4% 
\protect\pi$) as a function of Fermi energy $E_f$ (in units of t), for 
system size $N=24\times 24$ at $W=0.4t$ and $V_{so}=0.5t$. The dashed line 
with open triangles is SHC at periodic boundary condition averaged over $200$ 
disorder configurations. The solid line with filled triangles is SHC 
averaged over $10$ disorder configurations and $16\times 16$ different 
boundary phases.} 
\label{fig:fig2} 
\end{figure} 
 
The calculated $\sigma_{sH}$ at a weak disorder strength $W=0.4t$ with $200$ 
disorder configuration average, is shown in Fig.\ 1 (dashed line with open 
triangles) as a function of the Fermi energy $E_f$ with the system size $% 
N=24\times 24$ under periodic boundary condition (PBC) with $% 
\theta_x=\theta_y=0$. Clearly the fluctuations of the SHC are very strong 
much larger than $e/4 \pi$, in contrast to IQHE where the fluctuations in 
Hall conductance only appear between plateaus~\cite{fisher} with an 
amplitude smaller than $e^2/h$ (when converting to the units of spin 
conductance, $e^2/h $ is equal to $e/4\pi$). 
 
The SHC is also very sensitive to the boundary phases that we impose at the 
edges of the finite-size system. For example, antiperiodic boundary 
condition ($\theta_x=\theta_y=\pi$) will move the positive and negative 
peaks in Fig.\ 1 to different $E_f$'s. Remarkably, as we change $% 
\theta=(\theta_x,\theta_y)$ in the $2\pi\times 2\pi$ phase space, we find 
that the abnormal positive and negative fluctuations of the SHC at different 
$\theta$ can largely cancel each other. Thus $\langle\sigma_{sH}\rangle$ 
averaged over $16\times 16$ different $\theta$'s becomes a smooth function 
of $E_f$ with a value close to $0.3\frac {e} {4 \pi}$, which is also plotted 
in Fig.\ 1 as the line with the filled triangles. Completely similar 
behavior of $\sigma_{sH}$ upon boundary phase average has been observed for 
all $W$'s and $V_{so}$'s. 
 
\begin{figure}[tbp] 
\begin{center} 
\vskip-2.8cm \hspace*{-0.2cm} \includegraphics[width=2.2in]{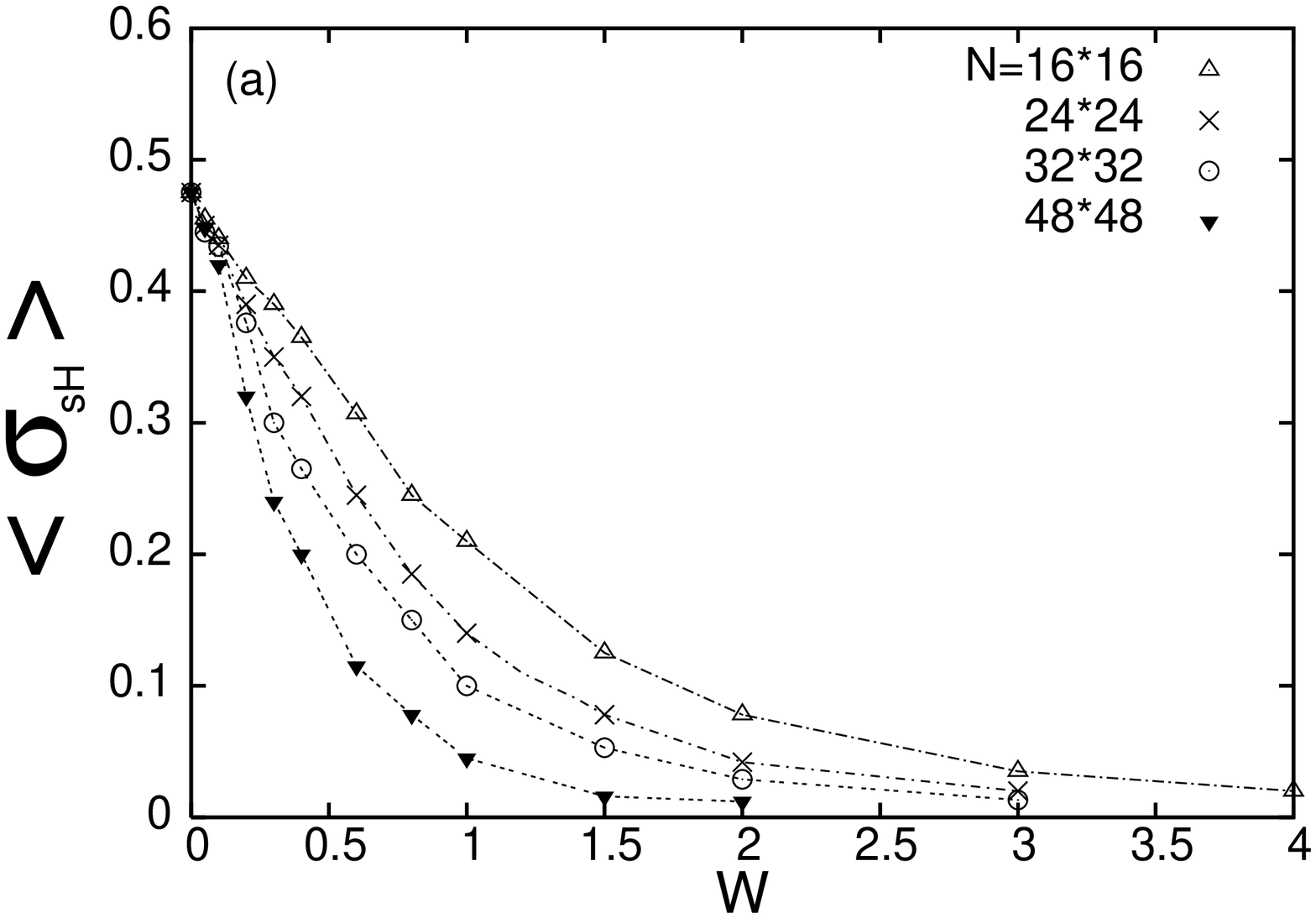} \vskip% 
-2mm 
\par 
\vskip-2.2cm \hspace*{-2.9cm} \includegraphics[width=2.14in]{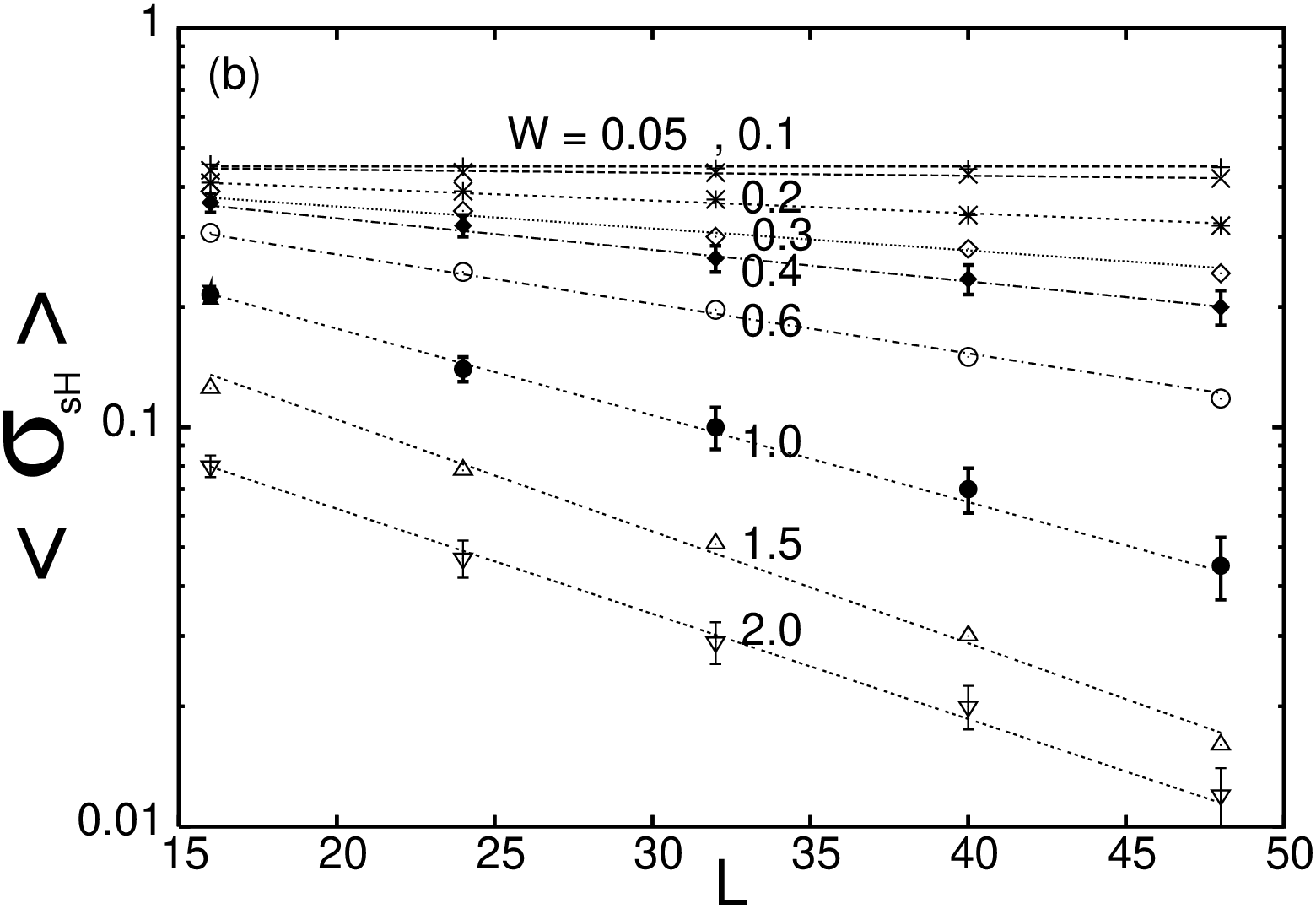} \vskip% 
-2mm 
\end{center} 
\par 
\vskip-0.9cm \hspace*{-0.02cm} 
\caption{ (a) The disorder and boundary phase averaged SHC $\langle\protect% 
\sigma_{sH}\rangle$ (in units of $e/4\protect\pi$) at $E_f=-3.75t$ as a 
function of disorder strength $W$ (in units of t) for sizes $N=16\times 
16,24\times 24,32\times32$ and $48\times 48$. (b) $<\protect\sigma_{sH}>$ 
for $W$ between $0.05$--$2.0$ $t$ as a function of sample length $% 
L=16,24,32,40$, $48$.} 
\label{fig:fig2} 
\end{figure} 
 
We note that the eigenstates of a finite-size system become 
boundary-phase independent only in the thermodynamic limit. 
However, when we apply an electric field through a time-dependent 
vector potential, it acts as a generalized boundary phase evolving 
with time.  Thus the SHC averaged over time is equivalent to the 
boundary phase averaged SHC. Similar boundary phase averaged 
charge Hall conductance gives rise  to a topological invariant 
Chern number \cite{thou, dns96}. In Fig.\ 2(a), we show 
$\langle\sigma_{sH}\rangle$ as a 
function of disorder strength $W$ (in units of $t$) for system sizes $% 
N=16\times 16, 24\times 24, 32\times 32, 48\times48$ with $V_{so}=0.5t$ at a 
fixed Fermi energy $E_f=-3.75t$ near the band bottom, where the spectrum of 
the lattice model approaches the continuum one. Each data point is averaged 
over random disorder configurations (about $200$ for $L=48$ and $500-1000$ 
for smaller sizes) and boundary phases ($16-576$ different $\theta$ for each 
disorder configuration). At weak $W$ limit, $\langle\sigma_{sH}\rangle$ 
reaches the value of 0.48$(\frac e {4 \pi})$ for all sizes, in agreement 
with the predicted intrinsic value for the pure system\cite{sinova}. 
 
The SHC decreases continuously with the increase of $W$ as shown in Fig. 
2(a). At very small $W=0.05t$, the SHC is essentially a constant close to 
the intrinsic value with negligible variation with $L$. However, the lack of 
a flat region in $\langle \sigma _{sH}\rangle $ at $W\rightarrow 0$ limit 
suggests that the effect of disorder is intrinsically important. Very 
similar results are obtained for other Fermi energies and $V_{so}$'s. 
Furthermore, we have also seen a monotonic decrease of $\langle \sigma 
_{sH}\rangle $ with increasing $L$ for $W\geq 0.1t$. To quantitatively 
understand such a behavior, we plot $\langle \sigma _{sH}\rangle $ in a 
logarithmic scale as a function of $L$ in Fig.\ 2(b) at different $W$'s with 
typical error bar shown. %Once we increase $W$ to $W=0.1t$, 
%a monotonic decrease of the SHC with $L$ becomes very clear, and 
In fact all the data can be nicely fitted into straight lines, as shown in 
Fig.\ 2(b), which indicates an exponential decay law 
\begin{equation} 
<\sigma _{sH}>=c_{0}\exp {(-L/\xi _{\mbox{\tiny S}})} 
\end{equation}% 
where the constant $c_{0}$ is insensitive to $W$ (for the whole range of $W$% 
's we find $c_{0}=0.46\pm 0.04$ except for the bottom curve with $W=2t$). 
Therefore, the scaling behavior of SHC can be fitted into the general 
one-parameter scaling theory\cite{local}: $\sigma _{sH}=f(L/\xi _{% 
\mbox{\tiny S}})$. Here, $\xi _{\mbox{\tiny S}}$ is a spin-related 
characteristic length scale determined by fitting the data in Fig.\ 2(b) 
into Eq. (3), which approximately follows a power-law behavior $\xi _{% 
\mbox{\tiny S}}=13.8/(W/t)^{1.5}$. 
It suggests a finite length scale for any 
weak disorder. Thus SHC remains finite with the sample size up to $\xi _{% 
\mbox{\tiny S}}$ (it can be as large as, say, a few hundreds of the lattice 
constant at weak disorder strength $W=0.1t$), and is extrapolated to zero 
exponentially in the thermodynamic limit. 
 
\begin{figure}[tbp] 
\begin{center} 
\vskip-3.0cm \hspace*{-1.9cm} \includegraphics[width=2.3in]{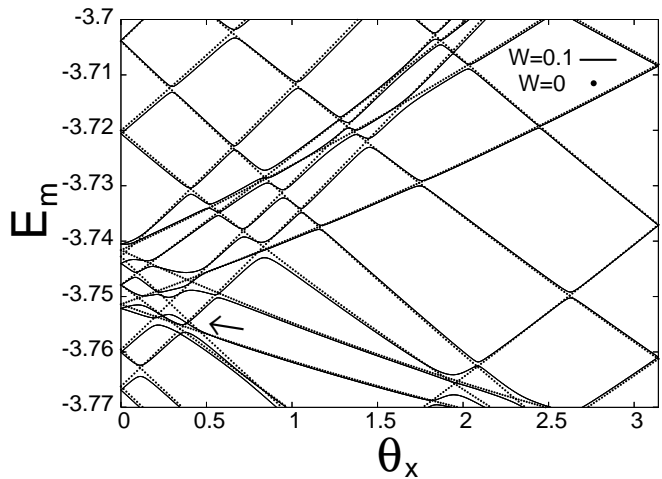} 
\vspace*{0.1cm} \vskip-0.6cm \hspace*{-0.6cm} 
\end{center} 
\caption{Energy levels of the 2D Rashba model (in units of $t$) with open 
boundary along $y$-axis as a function of the twisted phase $\protect\theta_x$ 
for a system of $V_{so}=0.1t$, $E_f=-3.75t$ and $N=32\times 32$. The 
solid-line is for $W=0.1t$ and the dotted-line is for $W=0$. The region 
indicated by the arrow is enlarged in Fig. 4(a).} 
\label{fig:fig2} 
\end{figure} 
 
\begin{figure}[tbp] 
\begin{center} 
\vspace*{2.6cm} \hspace*{3.cm} \includegraphics[width=2.6in]{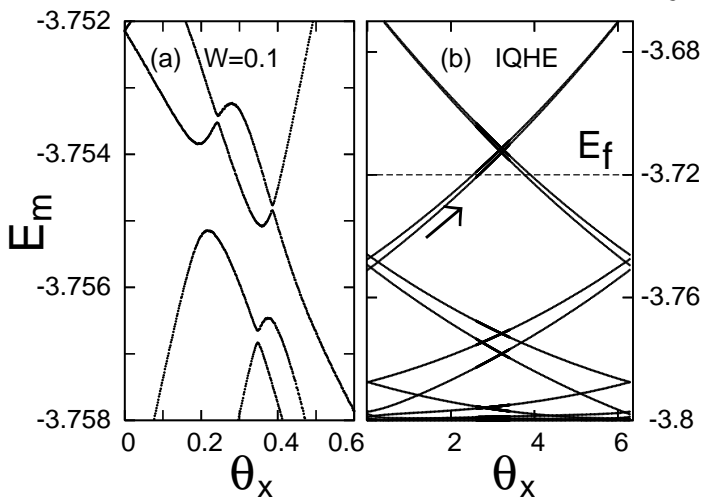} \vskip% 
-2.4cm \vskip-0.5cm 
\end{center} 
\caption{(a) Enlarged energy levels from Fig. 3 to show the avoided level 
crossing. (b) Robust level crossing in IQHE which leads to the adiabatic 
transfer of edge states. This is calculated for the same parameter used in 
Fig. 3 with an additional perpendicular magnetic field. } 
\label{fig:fig2} 
\end{figure} 
 
To further probe the spin transfer and accumulation associated with SHE, we 
perform numerically a Laughlin's gauge experiment\cite{laugh,halp} by 
adiabatically inserting flux to a system which is open along the $y$ 
direction and periodic  along the $x$ direction. This is the same geometry 
considered by Laughlin for IQHE and the adiabatic insertion of one flux 
quantum is equivalent to varying the twisted boundary phase $\theta_x$ by $% 
2\pi$ (note that electric field can also be related to the time dependent $% 
\theta _x$). By diagonalizing the Hamiltonian (1) under open boundary 
condition along the $y$ axis, at $2000$ different $\theta_x$'s, the 
resulting energies, $E_m$, around $E_f=-3.75t$ ($N=32\times 32$, $V_{so}=0.1t 
$ and $W=0.1t$) form solid lines shown in Fig. 3. Note that the energy 
spectrum is symmetric about $\theta_x=\pi$, thus only the half of the 
spectrum with $\theta_x \leq \pi $ is shown. A careful examination of these 
lines reveals that each $E_m$ goes up and down, making several large angle 
turns due to backward scatterings. These energy levels never cross with each 
other, except for at $\theta_x=0$ and $\pi$, where two levels become exactly 
degenerate (Kramers degeneracy). Nontrivial topological property\cite% 
{haldane,dns96} associated with Kramers degeneracy will give rise to 2D 
delocalization\cite{dns96}. The region pointed by an arrow near the left 
bottom corner of Fig. 3 is enlarged in Fig. 4(a), which clearly demonstrates 
the non-crossing feature. This is a general observation for all systems that 
we have checked with different $V_{so}$'s, $W$'s and $N$'s up to $N=200 
\times 200$, in accordance with the level-repulsion rule of the disordered 
system. Then if one follows any Kramers degenerated pair of states starting 
from $\theta_x=0$ to $\theta=2\pi$, one will always go back to exactly the 
same pair of states. Namely, after the adiabatic insertion of one flux 
quantum, all states evolve exactly back to the starting states. Thus it will 
result in no spin pumping between two edges and zero spin accumulation at 
the open boundary. By contrast, in the absence of disorder, all energy 
levels will simply evolve following the apparent straight lines (dotted 
lines in Fig. 3) and cross each other with the increase of $\theta_x$. Then 
after the insertion of one flux quantum, each energy level evolves into a 
new state which leads to spin transfer across the sample. But this is the 
trivial case of level crossing due to the absence of disorder scattering. 
Therefore, the topology of the states becomes completely different with the 
turn on of a weak disorder, although the energies themselves only shift by 
very small amounts as shown in Fig. 3. 
 
For comparison, we show a nontrivial case of level crossing in the 
IQHE for the same Hamiltonian (1) in the presence of a 
perpendicular magnetic field/flux ($\Phi_B= 2\pi/32$ per 
plaquette\cite{dns96}). As shown in Fig. 4(b), the energy levels 
do show simple crossing. This is due to the chiral symmetry of the 
edge states, which suppresses any backward scattering. Through 
inserting one flux quantum, the two occupied states below $E_f$ 
are pumped onto states above $E_f$ indicated by the arrow in Fig. 
4(b), which leads to two electrons transferred across from one 
edge to another, corresponding to the IQHE with $\sigma_H=2e^2/h$. 
We further note the existence of a topological SHE in an electron 
system of 2D graphene with Haldane's model\cite{haldane1} with 
SOC\cite{kane}, where quantized edge transfers are also 
detected. 
 
To summarize, we have shown that the  SHC persists in the 
presence of random disorder scattering for finite size systems up to a 
characteristic length scale beyond which it decreases and vanishes 
exponentially,  in agreement with  the analytic calculations involving vertex 
corrections~\cite{sh7,sh8}. Thus the length scale $\xi_s$ should also be 
identified as the spin-relaxation length defined in Ref.\onlinecite{sh8}.
  Furthermore, by performing numerically a 
Laughlin's gauge experiment\cite{laugh,halp}, we have found that all energy 
levels cannot cross each other and thus must result in zero spin transfer 
between the edges and zero spin accumulation at the boundary after the 
insertion of one flux quantum. This is in contrast to the quantized charge 
transfer in the QHE\cite{laugh,halp}, which is associated with the 
nontrivial topological property of quantum Hall systems\cite{arou,top,dns}. 
Therefore we establish the vanishing of the topological SHE in the 
disordered 2D Rashba model  beyond a linear response 
theory with the nonconserved electron spin. The backward scattering which 
reverses electron momentum and spin leading to zero spin transfer, is not 
properly taken into account in the linear response theory for SHC as it 
breaks current continuity condition. 
While finite SHC in four-terminal setup is obtained from LB formula\cite{sh10} 
by measuring  spin current inside the ideal leads, it is
related to boundary spin current in mesoscopic length
scale, which cannot be intepreted as bulk SHE due to the nonconservation
of the spin current.

\textbf{Note added:} After this paper was finished, we became 
aware of a numerical work in which the vanishing SHC in the 
thermodynamic limit for an electron Rashba model was also 
concluded.\cite{sh91} 
 
\textbf{Acknowledgment:} The authors would like to thank L. Balents, J. Hu, 
A. H. MacDonald, B. K. Nikolic, J. Sinova, M. Wu, and S. C. Zhang for 
stimulating discussions. This work is supported by ACS-PRF 41752-AC10, 
RC grant CC5643, the NSF grant/DMR-0307170 (DNS) and NSF 
MRSEC grant/DMR-0213706 at the Princeton Center for Complex 
Materials (FDMH), a grant from the Robert A. Welch Foundation (LS), and  
NSFC grants 10374058 and 90403016 (ZYW). 
DNS and FDMH wish to thank the KITP  
for support from NSF PHY99-07949.

\end{document}